\documentclass{pasj00}

\begin{document}
\SetRunningHead{Matsunaga et al.}
{Intracluster Dust of Globular Clusters}
\Received{2008/05/22}
\Accepted{2008/09/12}

\title{An AKARI Search for Intracluster Dust of Globular
Clusters\thanks{Based on observations with AKARI,
a JAXA project with the participation of ESA.}}

\author{Noriyuki \textsc{Matsunaga},\altaffilmark{1,2}
~Hiroyuki \textsc{Mito},\altaffilmark{3}
~Yoshikazu \textsc{Nakada},\altaffilmark{3,4}
~Hinako \textsc{Fukushi},\altaffilmark{4}
~Toshihiko \textsc{Tanab\'{e}},\altaffilmark{4} \\
~Yoshifusa \textsc{Ita},\altaffilmark{5,6}
~Hideyuki \textsc{Izumiura},\altaffilmark{7}
~Mikako \textsc{Matsuura},\altaffilmark{5,2}
~Toshiya \textsc{Ueta},\altaffilmark{8}
~and
Issei \textsc{Yamamura}\altaffilmark{6}
}
\altaffiltext{1}{Department of Astronomy, Kyoto University,
Kitashirakawa Oiwake-cho, Sakyo-ku, Kyoto, Kyoto 606-8502}
\email{matsunaga@kusastro.kyoto-u.ac.jp}
\altaffiltext{2}{Research Fellow of the Japan Society
for the Promotion of Science}
\altaffiltext{3}{Kiso Observatory, Institute of Astronomy,
School of Science, the University of Tokyo, Mitake,
Kiso, Nagano 397-0101}
\altaffiltext{4}{Institute of Astronomy, School of Science,
the University of Tokyo, 2-21-1 Osawa, Mitaka, Tokyo 181-0015}
\altaffiltext{5}{National Astronomical Observatory of Japan,
2-21-1 Osawa, Mitaka, Tokyo 181-8588}
\altaffiltext{6}{Institute of Space and Astronautical Science,
Japan Aerospace Exploration Agency, \\
Yoshinodai 3-1-1, Sagamihara, Kanagawa 229-8510}
\altaffiltext{7}{Okayama Astrophysical Observatory,
National Astronomical Observatory of Japan, \\
Kamogata, Asakuchi, Okayama 719-0232}
\altaffiltext{8}{Department of Physics and Astronomy,
University of Denver, Denver, CO 80208, USA}


\maketitle

\begin{abstract}
We report the observations of 12 globular clusters with
the Far-Infrared Surveyor (FIS) on-board AKARI infrared
satellite. Our goal is to search for emission from
the cold dust within clusters.
We detect diffuse emissions toward NGC~6402 and 2808,
but the IRAS 100~$\mu$m maps show the presence of
strong background radiation.
They are likely emitted from the galactic cirrus,
while we cannot rule out the possible association of
a bump of emission with the cluster in the case of NGC~6402.
We also detect 28 point-like sources mainly
in the WIDE-S images ($90~\micron$). At least several
of them are not associated with the clusters
but background galaxies based on some external catalogs.
We present the spectral energy distributions (SEDs)
by combining the near-and-mid infrared data obtained with
the Infrared Camera (IRC) if possible.
The SEDs suggest that most of the point sources
are background galaxies.
We find one candidate of the intracluster dust which
has no mid-infrared counterpart unlike the other
point-like sources,
although some features such as its point-like appearance
should be explained before we conclude its intracluster origin.
For most of the other clusters, we have confirmed
the lack of the intracluster dust.
We evaluate upper limits of the intracluster dust mass
to be between $10^{-5}$ and $10^{-3}$~M$_\odot$ depending on
the dust temperature. The lifetime of
the intracluster dust inferred from the upper limits
is shorter than 5~Myr ($T_{\mathrm{d}}=70$~K)
or 50~Myr (35~K). Such short lifetime indicates
some mechanism(s) are at work to remove the intracluster dust.
We also discuss its impact on the chemical evolution
of globular clusters.
\end{abstract}

\section{Introduction}

Red giants in globular clusters are known to have low initial
masses of about 1~M$_\odot$ considering their high ages
of 10--13~Gyr.
Observations revealed that some of the red giants have
the mass loss accompanied with dust formation
(e.g.\ \cite{Frogel-1988}, \cite{Matsunaga-2005}).
Circumstellar dust has been detected in on-going space missions
({\it Spitzer Space Telescope}, SST,
Boyer \etal\ \yearcite{Boyer-2006}, \yearcite{Boyer-2008},
\cite{Origlia-2007}; AKARI, \cite{Ita-2007}).
Terminal velocities of the stellar winds from red giants,
10--15 km~s$^{-1}$ \citep{McDonald-2007},
are lower than typical escape velocities from massive
globular clusters, $\sim$~30~km~s$^{-1}$ \citep{Gnedin-2002}.
For this reason, one would expect that the released materials are
trapped and accumulated within clusters in the halo region.
On the other hand, the intracluster matter is removed by ram pressure
when clusters cross the galactic plane.
It was expected that gas of the order of
10~$M _\odot$ exists in a cluster depending on
its size and the time since the last passage through
the galactic plane \citep{Tayler-1975}.

In spite of the early expectations, most previous attempts
failed to detect intracluster matter and placed upper limits
which were lower than the expected amounts.
The surveys
range from the ones of dust to those of different forms of gas
(dust, \cite{Hopwood-1999};
molecular gas, \cite{Smith-1995};
atomic gas, \cite{Faulkner-1991};
ionized gas, \cite{Knapp-1996}; and references therein).
Among the challenges to search for intracluster dust,
the most secure detection was found in the globular cluster M~15
(\cite{Knapp-1995}, \cite{Evans-2003}).
Recently, \citet{Boyer-2006} confirmed the extended emission
of the intracluster dust, which they named IR1a, with SST.
They also obtained its temperature ($\sim 70$K) and dust mass 
($9 \times 10^{-4}$~M$_\odot$) based on the spectral energy
distribution (SED). The temperature was found to
be higher than that of the interstellar dust
(\cite{Draine-1985}; \cite{Sodroski-1987}).
The obtained mass is still 4 times smaller than the expected mass
released via the mass loss
of red giants.
Intracluster gas has been also detected
in a couple of cases (e.g.\ \cite{Faulkner-1991}, \cite{Freire-2001}).

The deficiency of the intracluster dust suggests that
some mechanism(s) are at work to remove or destroy the dust even
in the halo region. The discrepancy between the expected amount of
the released
dust and the upper limit is larger than ten for
some clusters whose intracluster dust has not been detected
\citep{Knapp-1995}. It remains to be understood what mechanism(s)
are efficient to what extent. For example, \citet{Lynch-1990}
proposed sputtering of dust grains by hot halo gas to explain
the deficiency. On the other hand,
Hopwood \etal\ (\yearcite{Hopwood-1998}, \yearcite{Hopwood-1999})
rejected the sputtering as the mechanism responsible for the removal
of the intracluster dust.
\citet{Penny-1997} presented a list of possible mechanisms.
Very recently, \citet{Umbreit-2008} demonstrated that
stellar collisions can generate the kinetic energy to remove
the intracluster dust and gas.

The fate of the released matter within the clusters
may have an effect on evolution of globular clusters.
Although it was once considered that stars in a globular cluster are
chemically homogeneous, more and more studies revealed evidence
of inhomogeneity. Readers are referred to \citet{Smith-1987},
for example, for early investigations mainly about carbon and
nitrogen abundances. Recently, the existence of stars enriched
in helium have attracted a lot of attention (\cite{Norris-2004},
\cite{Lee-2005}). Many authors have argued that those enrichment
results from the matter released from intermediate-mass
asymptotic giant branch (AGB) stars (e.g.\ \cite{Tsujimoto-2007},
\cite{Ventura-2008}). \citet{Suda-2007} proposed that
low-mass AGB
stars
can also play
an important role in generating stars with
enriched helium contents via intracluster-matter accretion.
However, it is not well studied whether the ejecta
of AGB stars are actually 
retained within clusters
in the context of dynamics of intracluster gas and dust.
The ejecta may disappear before they are
used for the generation of enriched stars.
The connection to such evolutionary scenarios
strengthens the importance  to understand
how the intracluster matter decreases faster than expected
if it is true.

The Far-Infrared Surveyor (FIS; \cite{Kawada-2007})
on-board AKARI satellite \citep{Murakami-2007}
provides us with a new opportunity to investigate the intracluster dust.
The FIS can take much deeper images in the far infrared
with much higher angular resolution than before.
Observations in the far infrared have high sensitivities
for the cold dust, $\sim$30~K.
We used the FIS to observe a dozen of globular clusters
as a part of the Mission Program to investigate
the mass-loss phenomenon in evolved stars in globular clusters
and nearby dwarf galaxies (AGBGA; PI, Y. Nakada).

We describe the observation and data reduction
in section \ref{sec:Observation}.
As we will present, we detect a few extended emissions
and several point-like sources. Analyses of the extended emissions
and the point-like sources are presented in section \ref{sec:diffuse}
and \ref{sec:points}, respectively. We argue that
most of them are not associated with the clusters, while
one point-like source remains to be a candidate of
the dust within the cluster (section \ref{sec:candidate}).
In section \ref{sec:Discussion},
we present discussions on the mass of the intracluster dust
(for both the possible detection and upper limits)
and discuss the impact on evolution of globular clusters.
Section \ref{sec:Summary} summarizes this work.

\section{Observations and data reduction}\label{sec:Observation}

\subsection{FIS observation}

\begin{table*}
  \begin{minipage}{180mm}
  \caption{Observation log: the observed globular clusters,
  their galactic coordinates, observation IDs, and dates.
  We also list background fluctuations in the obtained images
  in the unit of MJy/str, $\sigma_{65}$ for N60,  $\sigma_{90}$
  for WIDE-S, $\sigma_{140}$ for WIDE-L, and $\sigma_{160}$ for N160.
  \label{tab:obs}}
  \begin{center}
    \begin{tabular}{ccrrccrrrr}
      \hline
Name  & Alias & RA ($^\circ$) & Dec ($^\circ$) & 
   ID & Date & 
   $\sigma_{65}$ & $\sigma_{90}$ & $\sigma_{140}$ & $\sigma_{160}$ \\
\hline
NGC~1261 &       & 48.0637 & $-55.2169$ & 
   1700006-001 & 2006 Dec 29
   & 0.33 & 0.10 & 0.41 & 0.62 \\
NGC~1851 &       & 78.5263 & $-40.0472$ & 
   1700029-001 & 2007 Mar 01
   & 0.27 & 0.09 & 0.41 & 0.56 \\
NGC~1904 &  M~79 & 81.0442 & $-24.5242$ & 
   1700008-001 & 2006 Sep 10
   & 0.20 & 0.08 & 0.36 & 0.56 \\
NGC~2808 &       & 138.0108 & $-64.8631$ & 
   1701003-001 & 2007 Jul 09
   & 0.31 & 0.15 & 0.52 & 0.65 \\
NGC~5024 &  M~53 & 198.2304 & $+18.1692$ & 
   1700011-001 & 2006 Dec 31
   & 0.21 & 0.08 & 0.34 & 0.54 \\
NGC~5139 & $\omega$~Cen & 201.6913 & $-47.4769$ & 
   1700055-001 &
   2007 Jan 29 & 0.31 & 0.13 & 0.41 & 0.68 \\
NGC~5272 &  M~3  & 205.5467 & $+28.3756$ & 
   1701006-001 & 2007 Jul 03
   & 0.24 & 0.08 & 0.34 & 0.52 \\
NGC~5634 &       & 217.4054 & $-5.9764$ & 
   1700086-001 & 2007 Jan 27
   & 0.23 & 0.09 & 0.38 & 0.51 \\
NGC~5904 &  M~5  & 229.6408 & $+2.0828$ & 
   1700013-001 & 2007 Feb 06
   & 0.21 & 0.07 & 0.35 & 0.52 \\
NGC~6205 &  M~13 & 250.4229 & $+36.4603$ & 
   1701019-001 & 2007 Aug 22
   & 0.26 & 0.08 & 0.37 & 0.56 \\
NGC~6341 &  M~92 & 259.2804 & $+43.1364$ & 
   1700018-001 & 2007 Mar 01
   & 0.22 & 0.08 & 0.37 & 0.54 \\
NGC~6402 &  M~14 & 264.4004 & $-3.2458$ & 
   1700097-001 & 2006 Sep 16
   & 0.25 & 0.25 & 0.90 & 0.65 \\
\hline
    \end{tabular}
  \end{center}
  \end{minipage}
\end{table*}

We observed 12 globular clusters by using the AKARI/FIS
between 2006 May and 2007 August (table \ref{tab:obs}).
The FIS has four sets of array and filter, named
N60, WIDE-S, WIDE-L, and N160.
The central wavelengths of their responses are approximately
65, 90, 140, and $160\micron$. The formats of the arrays are
long and thin, e.g.\ 3$\times$20 for WIDE-S, and they are designed
to make scans along ecliptic meridian.
Our data were obtained under the FIS01 slow-scan observational mode
with a scan rate of eight arc-seconds per second
and a reset interval of two seconds.
In this mode, the arrays make two sets of
round-trip scans, which produces maps with the fields-of-view
of about $20\arcmin \times 8\arcmin$. The long side of the field
is aligned in the direction of ecliptic meridian.
Thus, we obtained images around the globular clusters
in four filters between 65 and $160~\micron$.

\subsection{Data reduction}

\begin{figure*}
\begin{minipage}{160mm}
  \begin{center}
    \FigureFile(146m,200mm){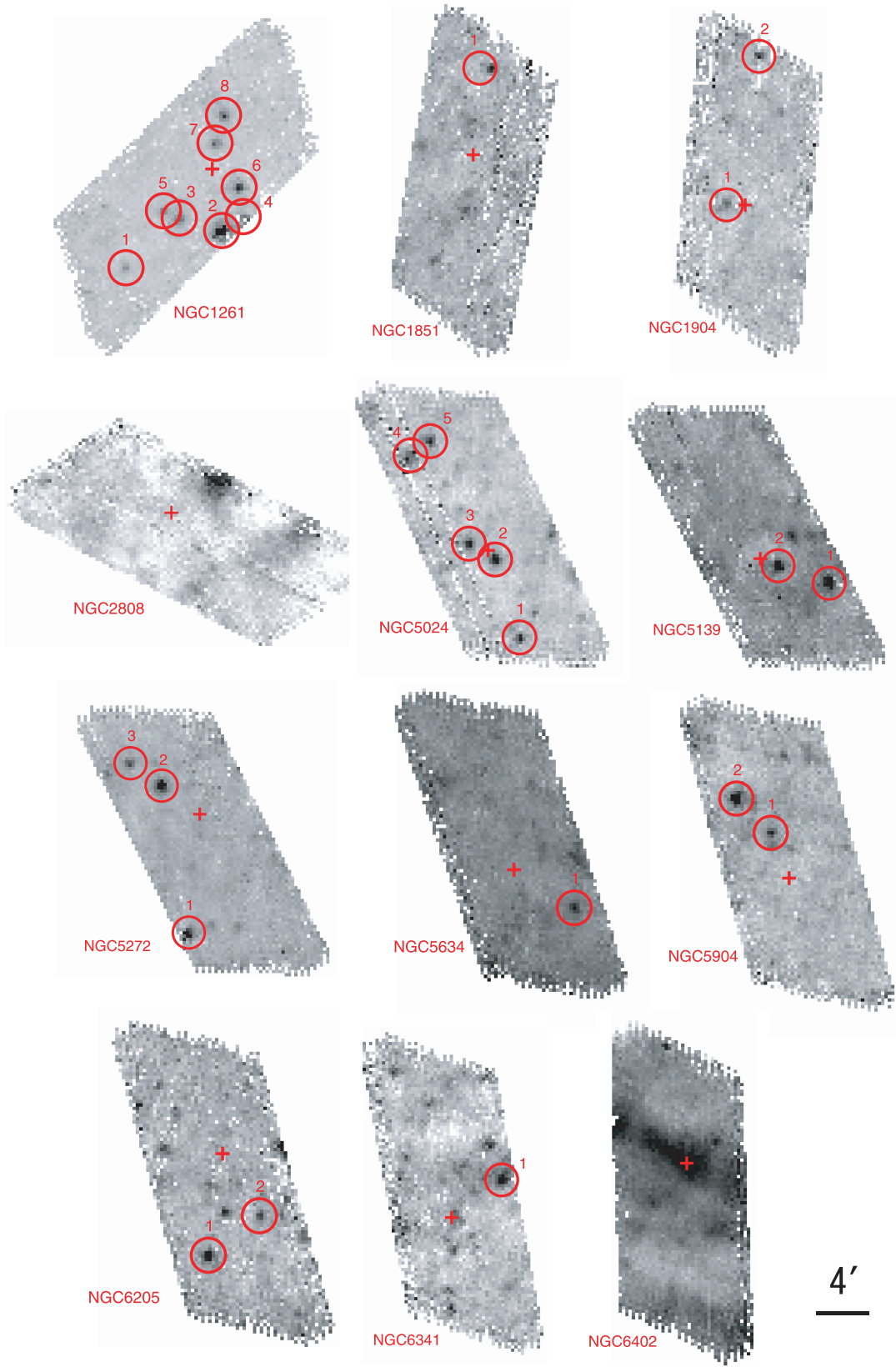}
  \end{center}
\caption{WIDE-S images of the observed fields.
Each image has a field-of-view of about $20\arcmin \times 8\arcmin$.
The long side of the field
is aligned in the direction of ecliptic meridian.
The north in the equatorial system is up, and the east is left.
Cross symbols indicate the centers of the globular clusters.
The detected
point sources are indicated by circles with the identification
numbers.
\label{fig:charts}}
\end{minipage}
\end{figure*}

The FIS data were first processed with an official data-analysis
tool for the slow-scan observation (FIS Slow-Scan Toolkit
version 20070914).
FITS-format images were created after several procedures
such as combination of scanned signals
into two-dimensional image and flux calibration.
The pixel size of the image was chosen to be 15" and 30"
for $65/90~\micron$ and $140/160~\micron$ respectively.
Each pixel value was obtained by taking a median of
typically five to ten measurements from the round-trip scans.
We present the obtained images in the WIDE-S filter
in figure \ref{fig:charts}.

We can find several point sources with expected point spread
functions, FWHM $\sim 40\arcsec$ in WIDE-S. Diffuse emissions
are also visible in the images of NGC~2808 and NGC~5024.
Most of the sources we detected are visible in the WIDE-S images
while some of them are also detected in the N60 images.
Only a couple of point sources are found in the WIDE-L images,
and none in the N160 images. We will focus on the results
in the images in two shorter wavelengths unless otherwise mentioned.
For further analysis,
we used the IRAF\footnote{IRAF is distributed by
the National Optical Astronomy Observatories, which are operated
by the Association of Universities for Research in Astronomy,
Inc., under cooperative agreement with the National Science
Foundation.} as we present in the following sections.

\subsection{Background fluctuation}\label{sec:skysigma}

In order to estimate background fluctuations of the images,
a histogram of pixel values for each image was created and 
the peak corresponding to the background was fitted with a Gaussian
distribution. The standard deviation $\sigma$ of
the fitted Gaussian is adopted as the fluctuation of the background,
and listed in table \ref{tab:obs}. In WIDE-S, for example,
most of the images have the fluctuation of about
$\sigma \sim 0.08$ [MJy/str].
We find the largest fluctuation in the NGC~6402 image.
This is caused by the high background emission of
the galactic cirrus \citep{Sodroski-1987}.
The images for NGC~2808 and NGC~5139
also have slightly
larger dispersions. These three clusters have the locations
closest to the galactic plane.

\section{Analysis of diffuse emissions}\label{sec:diffuse}

\subsection{NGC~6402}
\label{sec:NGC6402}

\begin{figure}
\begin{minipage}{80mm}
  \begin{center}
    \FigureFile(60mm,60mm){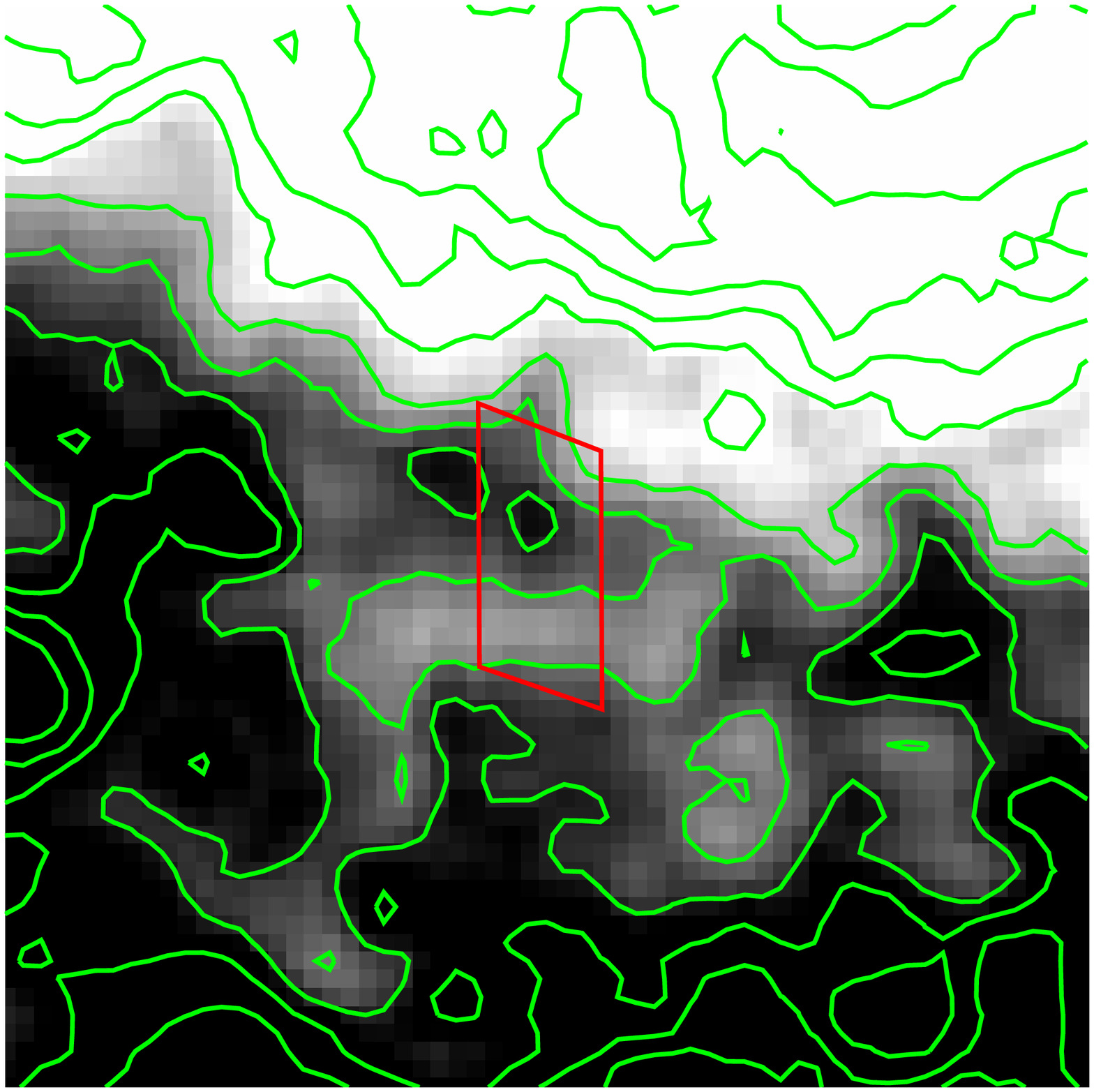}
  \end{center}
\caption{
IRAS $100~\micron$ map of 1.5 degrees square around NGC~6402.
The north is up and the east is left.
The field observed in our AKARI pointed observation
in the WIDE-S filter is indicated by a red parallelogram.
The contours of the IRAS $100~\micron$ flux density are overplotted
at the levels of 15, 16, 17, $\cdots$, 25 [MJy/str].
\label{fig:NGC6402IRAS}}
\end{minipage}
\end{figure}

The image of NGC~6402 (figure \ref{fig:charts}) reveals
extended emission.
The cluster center is well projected to within the bump
although the bump of emission does not have a sharp peak.
In figure \ref{fig:NGC6402IRAS}, we present the IRAS $100~\micron$ map
of 1.5 degrees square around NGC~6402.
The field of view of the FIS is indicated by a red polygon.
Strong large-scale emission clearly exists as a background around this region.
In the WIDE-S $90~\micron$ image, for example,
the bottom of the background emission is about 15 MJy/str
and the emission reaches about 16 MJy/str around the cluster center.
The observed bump of emission is
comparable to the fluctuation of the large-scale emission.
This cluster is the one closest to
the galactic plane among our samples
($l=+21.324 ^{\circ}, ~b=+14.804 ^{\circ}$).
There are three possible origins of the bump of emission
around NGC~6402. The first one is that the whole emission
comes from the interstellar dust (or galactic cirrus)
and is not associated with the cluster.
Alternatively, the emission can arise from from the intracluster dust
within NGC~6402.
The last possibility is that the nearby interstellar dust
is heated by the cluster.

\begin{figure*}
\begin{minipage}{140mm}
  \begin{center}
    \FigureFile(120mm,100mm){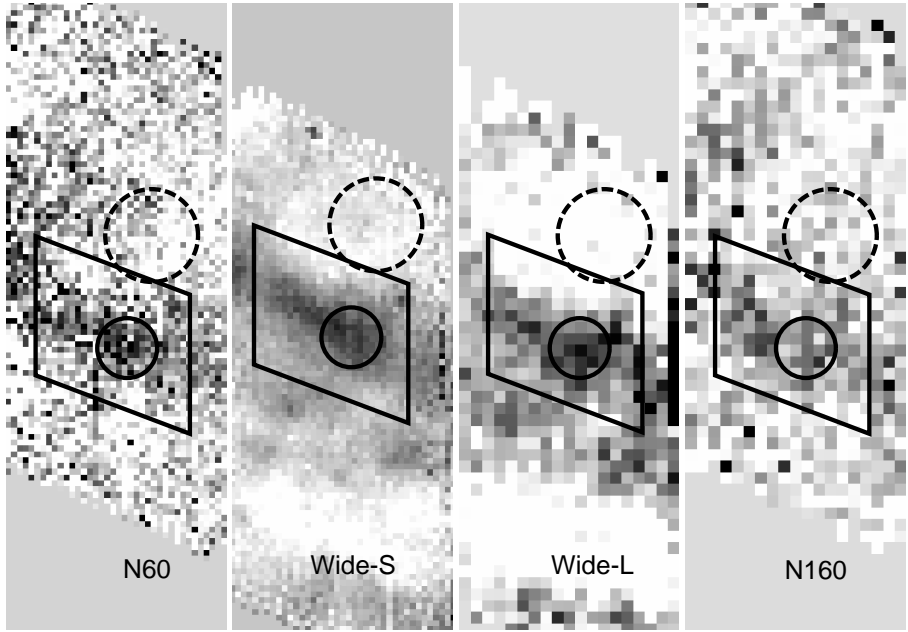}
  \end{center}
\caption{
Four-band images of NGC~6402 and apertures for photometry.
The thick circle and the polygon were used to measure the extended
emission while the dashed circle was used to estimate
the large-scale background emission.
\label{fig:N6402apertures}}
\end{minipage}
\end{figure*}

\begin{figure}
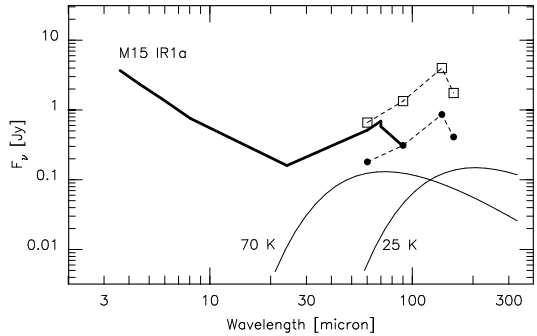

\begin{minipage}{80mm}
  \begin{center}
    \FigureFile(70mm,50mm){fig4.ps}
  \end{center}
\caption{
Spectral energy distribution of the enhancement of
the diffuse emission at around NGC~6402.
The ones with different apertures are plotted with different symbols
(filled circles, a circle aperture within $r\leq r_{\mathrm{h}}$;
open squares, a polygon aperture).
We also plot the blackbody radiation with 25~K and 70~K
(thin curves)
and the SED of the stellar and dust components near the core of M~15
reported by \citet{Boyer-2006} (thick curve).
\label{fig:SED6402}}
\end{minipage}
\end{figure}

In order to investigate the nature of the diffuse emission,
we performed photometry of the emission around the cluster.
First, we estimated the background emission within
an aperture indicated by the dashed circle, which we call sky region,
in figure \ref{fig:N6402apertures}. Flux densities are at minimum
around the sky region in the FIS images.
Then, fluxes above the background level
of the sky region were obtained with two different apertures:
a circle with the half-mass radius
of the cluster around its center ($r < r_h = 1.29\arcmin$) and
a polygon with an area of $3.43\times 10^{-6}$ str as shown in
figure \ref{fig:N6402apertures}.
The obtained fluxes are listed in table \ref{tab:ExtendedEmission}.
Figure \ref{fig:SED6402} plots the fluxes
altogether with the blackbody radiation of 70~K and 25~K.
The dust temperature $T_{\mathrm{d}}$ of the bump detected
toward NGC~6402 is similar to that of the cirrus
in the galactic plane ($\sim 25$~K; \cite{Sodroski-1987}).
We also plot the SED of the
stellar and dust components, $T_{\mathrm{d}}=70$~K,
near the core of M~15 reported by \citet{Boyer-2006}.
The SED of the dust component is clearly different
from the case of NGC~6402. However, the temperature
of a possible intracluster dust cloud
can be uncertain because of stochastic effects
of the UV flux from hot post-AGB stars,
interacting binaries and millisecond pulsars.
Millisecond pulsars and X-ray binaries have been found
in M~15 (\cite{Camilo-2005}, \cite{Dieball-2005},
and references therein).
There are no such exotic objects found in NGC~6402
although the surveys may not be complete.

In conclusion, the strong background suggests
that the bump of emission toward NGC~6402 is likely to be
the fluctuation of the galactic cirrus,
but we cannot rule out the possible presence of the dust associated
with the cluster.
It may be interesting to compare the elongation of the bump toward
west-northwest with the proper motion of the cluster
which is not yet known.

\begin{table}
\begin{minipage}{80mm}
\caption{
Fluxes of the diffuse emissions around NGC~6402 and NGC~2808.
The apertures are indicated in figure \ref{fig:N6402apertures}
and \ref{fig:N2808apertures}. 
\label{tab:ExtendedEmission}}
\begin{center}
\begin{tabular}{lrrrrr}
\hline
Aperture & \multicolumn{1}{c}{Area} & $F_{65}$ & $F_{90}$ &
   $F_{140}$ & $F_{160}$ \\
&  \multicolumn{1}{c}{$10^{-7}$~str} & \multicolumn{1}{c}{Jy} &
   \multicolumn{1}{c}{Jy} & \multicolumn{1}{c}{Jy} &
   \multicolumn{1}{c}{Jy} \\
\hline
\multicolumn{6}{c}{NGC~6402 (figure \ref{fig:N6402apertures})} \\
Filled circle$^\dagger$ & 4.5 & 0.18 & 0.31 & 0.86 & 0.41  \\
Polygon$^\dagger$ & 34.3 & 0.66 & 1.35 & 3.98  & 1.75  \\
Dashed circle & 10.7 & 16.4 & 16.3 & 29.6 & 23.9  \\
\hline
\multicolumn{6}{c}{NGC~2808 (figure \ref{fig:N2808apertures})} \\
Polygon$^\dagger$ & 6.2 & \multicolumn{1}{c}{---} & 0.31 & 0.82 &
   \multicolumn{1}{c}{---}  \\
Dashed circle & 10.7 & \multicolumn{1}{c}{---} & 9.4 & 19.1 &
   \multicolumn{1}{c}{---}  \\
\hline
\end{tabular}
\end{center}
$^\dagger$ We subtracted the background level estimated
in the corresponding dashed circle for each cluster.
\end{minipage}
\end{table}

\subsection{NGC~2808}

\begin{figure}
\begin{minipage}{80mm}
  \begin{center}
    \FigureFile(60mm,60mm){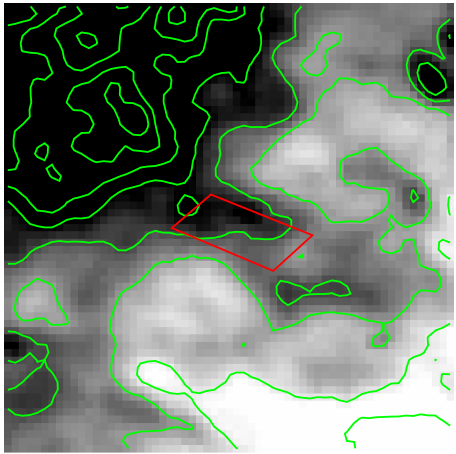}
  \end{center}
\caption{
IRAS $100~\micron$ map of 1.5 degrees square around NGC~2808.
The north is up and the east is left.
The field observed in our AKARI pointed observation
in the WIDE-S filter is indicated by a red parallelogram.
The contours of the IRAS $100~\micron$ flux density are overplotted
at the levels of 8, 9, 10, $\cdots$, 16 [MJy/str].
\label{fig:NGC2808IRAS}}
\end{minipage}
\end{figure}

\begin{figure}
\begin{minipage}{80mm}
  \begin{center}
    \FigureFile(70mm,70mm){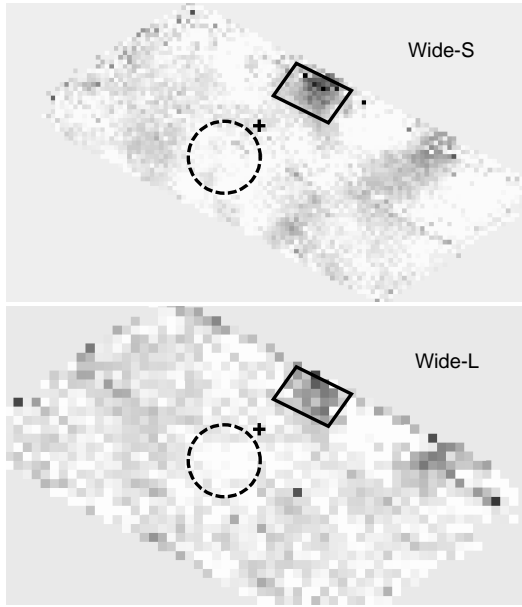}
  \end{center}
\caption{
Two-band images of NGC~2808 and apertures for photometry.
The thick polygon was used to measure the extended emission
while the dashed circle was used to estimate
the large-scale background level.
\label{fig:N2808apertures}}
\end{minipage}
\end{figure}

The image for NGC~2808 also shows extended emissions
though the bumps are separated from the central coordinate of
the cluster (figure \ref{fig:charts}).
Figure \ref{fig:NGC2808IRAS} presents the IRAS $100~\micron$ map
of 1.5 degrees square around NGC~2808,
which has a large-scale fluctuation
as well as NGC~6402. We made photometry on the strongest bump
at around the edge with an polygon aperture indicated in
figure \ref{fig:N2808apertures}. The emissions are
clear only in the images of WIDE-S and WIDE-L, so that we
estimated the fluxes in the two filters and list them
in table \ref{tab:ExtendedEmission}. As we did in the previous
section, the background level is evaluated within the dashed circle
of two arc-minutes radius indicated
in figure \ref{fig:N2808apertures}. The ratio of $F_{90}/F_{140}$
is apparently similar to those obtained for NGC~6402 rather than
that of M~15 IR1a. The large-scale fluctuation in the IRAS
$100~\micron$ map is larger than the bump, $\lesssim$~1~MJy/str,
in the WIDE-S image.
Therefore, the diffuse emission in direction toward NGC~2808
can be entirely ascribed to the galactic cirrus.
The positional difference strongly supports that
the emission is not associated with the cluster.

\section{Analysis of point-like sources}\label{sec:points}

\subsection{Source list}

As can be seen in figure \ref{fig:charts},
several point-like sources are visible in the images.
We performed source detections and aperture photometry by using
IRAF/APPHOT package. First, the {\it daofind} task was used to detect
the sources with detection limits of 5~$\sigma$, where $\sigma$
was chosen to be the background fluctuations listed
in table \ref{tab:obs}. Each candidate was examined by eye
in order to exclude spurious pixels.
Although some fainter objects seem to be visible,
the number of spurious detections get large with a lower detection limit,
e.g. 3~$\sigma$. We decided to set the detection threshold of 5~$\sigma$.
The obtained source list was given to the {\it phot} task with
the apertures ($37.5\arcsec$) and sky annuli
($2.\arcmin25$--$3.\arcmin25$) listed in the FIS Data User
Manual\footnote{The manuals for both FIS and IRC are available at\\
http://www.ir.isas.jaxa.jp/ASTRO-F/Observation/}
(Version 1.3; \cite{Verdugo-2007}).
Flux calibration was performed based on the conversion factors,
also in the Manual.
No color correction was applied.

In table \ref{tab:objects}, we compile the catalog of
the point-like sources.
We have detected 28 objects in total.
Most of the detections above the 5-$\sigma$ limit are found in
WIDE-S images, except N6205 FIS3 detected only in the N60 image
which locates out of the WIDE-S field of view.
They are indicated by circles in figure \ref{fig:charts}.
The position accuracy is expected to be as good as $\pm~10\arcsec$.
The dominant source of errors in the obtained fluxes is
the uncertainty of the the conversion factors for the calibration
(20\%, \cite{Verdugo-2007}).

\begin{table}
\begin{minipage}{80mm}
\caption{
Point-like sources found in the direction toward globular clusters.
Right Ascensions, Declinations, and flux densities at $90~\micron$
(Wide-S) and $65~\micron$ (N60) are listed.
The last column indicates that the source was observed with
the IRC (T) or not (F). N5024~FIS2 is given the N flag since
it locates within the IRC fields of view but was not detected
(see section \ref{sec:candidate}).
\label{tab:objects}}
\begin{center}
\begin{tabular}{crrrrc}
\hline
ID & \multicolumn{1}{c}{RA} & \multicolumn{1}{c}{Dec} & $F_{90}$ &
   $F_{65}$ & IRC \\
& \multicolumn{1}{c}{deg} & \multicolumn{1}{c}{deg} & mJy & mJy \\
\hline
N1261~FIS1 & $48.248$ & $-55.336$ & $62$ & --- & F \\
N1261~FIS2 & $48.044$ & $-55.291$ & $413$ & $304$ & T \\
N1261~FIS3 & $48.134$ & $-55.276$ & $155$ & --- & T \\
N1261~FIS4 & $47.998$ & $-55.275$ & $108$ & $252$ & T \\
N1261~FIS5 & $48.168$ & $-55.267$ & $99$ & --- & T \\
N1261~FIS6 & $48.007$ & $-55.239$ & $187$ & $108$ & T \\
N1261~FIS7 & $48.058$ & $-55.185$ & $119$ & $27$ & T \\
N1261~FIS8 & $48.040$ & $-55.152$ & $150$ & $115$ & T \\
N1851~FIS1 & $78.515$ & $-39.936$ & $68$ & $91$ & F \\
N1904~FIS1 & $81.066$ & $-24.524$ & $57$ & --- & F \\
N1904~FIS2 & $81.020$ & $-24.332$ & $53$ & --- & F \\
N5024~FIS1 & $198.189$ & $18.060$ & $79$ & $79$ & F \\
N5024~FIS2 & $198.221$ & $18.158$ & $114$ & $86$ & N \\
N5024~FIS3 & $198.256$ & $18.177$ & $98$ & $94$ & T \\
N5024~FIS4 & $198.333$ & $18.288$ & $106$ & --- & F \\
N5024~FIS5 & $198.307$ & $18.306$ & $86$ & --- & F \\
N5139~FIS1 & $201.566$ & $-47.506$ & $195$ & $134$ & T \\
N5139~FIS2 & $201.663$ & $-47.484$ & $132$ & $91$ & T \\
N5272~FIS1 & $205.563$ & $28.221$ & $169$ & --- & F \\
N5272~FIS2 & $205.603$ & $28.413$ & $239$ & $301$ & T \\
N5272~FIS3 & $205.650$ & $28.442$ & $89$ & $52$ & F \\
N5634~FIS1 & $217.331$ & $-6.023$ & $66$ & --- & F \\
N5904~FIS1 & $229.664$ & $2.139$ & $55$ & --- & F \\
N5904~FIS2 & $229.706$ & $2.182$ & $150$ & $101$ & F \\
N6205~FIS1 & $250.444$ & $36.337$ & $99$ & --- & F \\
N6205~FIS2 & $250.367$ & $36.385$ & $42$ & --- & F \\
N6205~FIS3 & $250.536$ & $36.683$ & --- & $143$ & F \\
N6341~FIS1 & $259.192$ & $43.185$ & $95$ & --- & T \\
\hline
\end{tabular}
\end{center}
\end{minipage}
\end{table}

\subsection{Near-to-mid infrared magnitudes}

In order to study the natures of the FIS sources, 
we examined near- and mid-infrared images obtained
by using the Infrared Camera (IRC; \cite{Onaka-2007}) on-board
AKARI. Among the 12 clusters in our sample,
eight globular clusters were also observed with the IRC, namely
NGC~1261, NGC~1851, NGC~2808, NGC~5024, NGC~5139, NGC~5272, NGC~6205,
and NGC~6341. The IRC has three channels: NIR (N3 and N4),
MIR-S (S7 and S11), and MIR-L (L15 and L24). In the parentheses
given are the names of the filters we used with
the IRC02 observational mode. While NIR and MIR-S channels have
the same field of view, MIR-L points to a different
field. Thus, two pointing chances are necessary to obtain images 
of a certain field in
all the six filters (once in NIR and MIR-S at the same time, and
in MIR-L during another pointing). We obtained the six images
for the clusters listed above except NGC~5272 which was only observed
in the MIR-L filters.

Raw data were processed with the IRC imaging data pipeline,
version 070912 \citep{Lorente-2007}. Each IRC image has
a field of view of about 10-arcminutes-square.
around the center of a cluster. With the different shapes
of the observed fields,
some of the FIS sources in table \ref{tab:objects} are not located
within the IRC images. For those within the IRC frames,
most of the FIS sources
have the counterparts which are apparently red, i.e. they are bright
in longer wavelength (MIR-L) and faint in shorter wavelength (NIR)
compared with numerous normal stars. In a few cases such as
N5024 FIS3, there are more than one NIR objects within
the positional accuracy of the FIS source.
Nonetheless, it is not difficult to identify
a counterpart in the longer wavelength because we can find one-to-one
correspondence of the FIS objects to the ones in MIR-L images.
It is then possible to select the specific counterparts
in the near-infrared images in comparison with the MIR-L images
which have higher positional accuracies and resolutions than
the FIS images. For N5024 FIS2, we did not find the counterparts
even in the MIR-L images.

We used IRAF/DAOPHOT package for photometric measurements,
and the fluxes for
the FIS counterparts were calibrated into Jy scale,
based on the flux conversion factors released on 12 November 2007
(Tanab\'{e}, in preparation).
They were further transformed into magnitude scale based
on the fluxes of the 0-th magnitude:
343, 184, 75.0, 38.3, 16.0, and 8.05~Jy 
in N3, N4, S7, S11, L15, and L24 filters, respectively.
The obtained magnitudes are listed in table
\ref{tab:objects2}.

\subsection{Spectral energy distributions}

\begin{figure*}
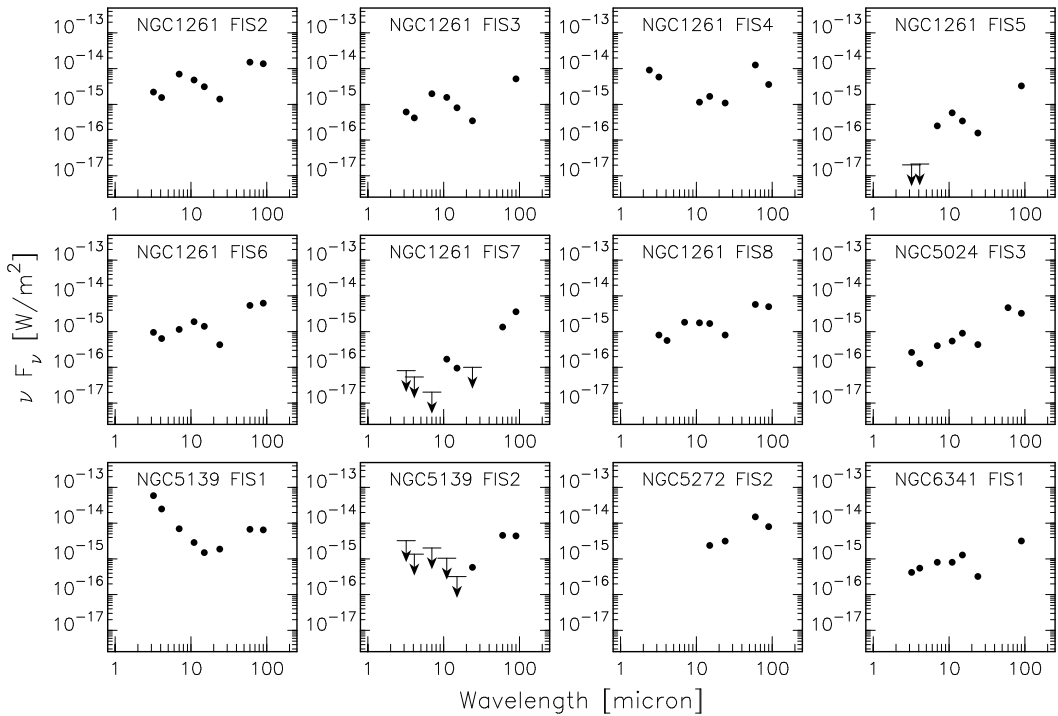

\begin{minipage}{160mm}
  \begin{center}
    \FigureFile(140mm,90mm){fig7.ps}
  \end{center}
\caption{
Spectral energy distributions of the objects observed both in
the FIS and the IRC. Arrows are indicated when no counterpart
was identified although the image includes the coordinate.
\label{fig:FISIRCSEDs}}
\end{minipage}
\end{figure*}

Combining the fluxes obtained with the FIS and IRC, we plot the SEDs
of the FIS point sources in figure \ref{fig:FISIRCSEDs}.
Many of them show a similar character: they have strong emissions
in the far infrared and there are bumps at around $10~\micron$.
These SEDs are very different from those of any stellar objects.
For example, the flux ratio between $100~\micron$ and $25~\micron$
ranges from 20 to more than 100, which is larger than
the observed values for post-AGB stars (0.1--4; \cite{Volk-1989})
or planetary nebulae (0.1--10; \cite{Pottasch-1984}).
Overall shape of the SED cannot be attributed to a single component
of blackbody radiation either.
Even if we assume that there are two components from
the cold intracluster dust and one or a group of star(s),
their SEDs in the near-to-mid infrared are unlike those
of normal stellar components in globular clusters
(compare with the SED of M15~IR1a in figure \ref{fig:SED6402}).

\begin{table*}
\begin{minipage}{155mm}
\caption{
Near-to-mid infrared magnitudes of the FIS objects.
Also listed are coordinates obtained in the IRC images,
in S11, whenever available, and differences from the coordinates
obtained in the FIS images ($\Delta r$).
\label{tab:objects2}}
\begin{center}
\begin{tabular}{crrrrrrrrr}
\hline
 \multicolumn{1}{c}{ID} & \multicolumn{1}{c}{RA} & \multicolumn{1}{c}{Dec} & \multicolumn{1}{c}{$\Delta r$} & \multicolumn{1}{c}{N3} & \multicolumn{1}{c}{N4} & \multicolumn{1}{c}{S7} & \multicolumn{1}{c}{S11} & \multicolumn{1}{c}{L15} & \multicolumn{1}{c}{L24} \\
 & \multicolumn{1}{c}{deg} & \multicolumn{1}{c}{deg} & \multicolumn{1}{c}{sec} & \multicolumn{1}{c}{mag} & \multicolumn{1}{c}{mag} & \multicolumn{1}{c}{mag} & \multicolumn{1}{c}{mag} & \multicolumn{1}{c}{mag} & \multicolumn{1}{c}{mag} \\
\hline
N1261 FIS2 & $ 48.0379$ & $-55.2906$ &  12 & 12.91 & 12.35 & 9.15 & 8.34 & 7.53 & 7.14 \\
N1261 FIS3 & $ 48.1286$ & $-55.2747$ &  12 & 14.30 & 13.77 & 10.52 & 9.56 & 9.00 & 8.66 \\
N1261 FIS4 & $ 47.9871$ & $-55.2795$ &  29 & 12.31 & 11.86 & \multicolumn{1}{c}{---~$^b$} & 9.89 & 8.21 & 7.41 \\
N1261 FIS5 & $ 48.1629$ & $-55.2657$ &  11 & \multicolumn{1}{c}{---~$^c$} & \multicolumn{1}{c}{---~$^c$} & 12.77 & 10.64 & 9.93 & 9.51 \\
N1261 FIS6 & $ 47.9992$ & $-55.2373$ &  17 & 13.82 & 13.31 & 11.11 & 9.35 & 8.39 & 8.42 \\
N1261 FIS7 & $ 48.0500$ & $-55.1835$ &  16 & \multicolumn{1}{c}{---~$^c$} & \multicolumn{1}{c}{---~$^c$} & \multicolumn{1}{c}{---~$^c$} & 11.97 & 11.32 & \multicolumn{1}{c}{---~$^c$} \\
N1261 FIS8 & $ 48.0321$ & $-55.1489$ &  19 & 14.01 & 13.45 & 10.61 & 9.43 & 8.20 & 7.74 \\
N5024 FIS3 & $198.2525$ & $+18.1797$ &  14 & 15.22 & 15.06 & 12.25 & 10.71 & 8.88 & 8.41 \\
N5139 FIS1 & $201.5630$ & $-47.5069$ &   8 & 9.35 & 9.34 & 9.16 & 8.90 & 8.33 & 6.82 \\
N5139 FIS2 & $201.6596$ & $-47.4846$ &   8 & \multicolumn{1}{c}{---~$^c$} & \multicolumn{1}{c}{---~$^c$} & \multicolumn{1}{c}{---~$^c$} & \multicolumn{1}{c}{---~$^c$} & \multicolumn{1}{c}{---~$^c$} & 8.10 \\
N5272 FIS2 & $205.6028$ & $+28.4147$ &   7 & \multicolumn{1}{c}{---~$^a$} & \multicolumn{1}{c}{---~$^a$} & \multicolumn{1}{c}{---~$^a$} & \multicolumn{1}{c}{---~$^a$} & 7.83 & 6.26 \\
N6341 FIS1 & $259.1917$ & $+43.1881$ &  10 & 14.71 & 13.48 & 11.51 & 10.28 & 8.49 & 8.74 \\
\hline
\end{tabular}
\end{center}
$^a$ No image was taken in this filter.\\
$^b$ The object located out of the field of view.\\
$^c$ No counterpart was detected in this filter.\\
\end{minipage}
\end{table*}

The SEDs in figure \ref{fig:FISIRCSEDs} resemble those of galaxies
(see Fig. 4 in \cite{Sajina-2006}, for example).
\citet{Pearson-2007} calculated the expected AKARI colors
for various kinds of galaxies. The flux ratio $F$(S7)/$F$(WIDE-S)
ranges from 0.001 to 0.1, which agrees with the range of our sample,
0.006--0.85.
The bumps at around $10~\micron$ can be attributed to
the emission of polycyclic aromatic hydrocarbon
(\cite{Lee-2007}, \cite{Sakon-2007}).

N5139 FIS1 has a different SED compared with others. This source
is identical to the $70~\micron$ source SSTOCEN J132615.23-473024.56
(Source \#4) detected by \citet{Boyer-2008}.
As they mentioned, it is potentially affected by blending
between a neighboring bright star and a fainter real counterpart
of the far-infrared emission. So that, it may be inappropriate to
discuss its nature based on the obtained SED.

\subsection{Identifications with background galaxies}
\label{sec:background}

\begin{figure}
\begin{minipage}{80mm}
  \begin{center}
    \FigureFile(65mm,50mm){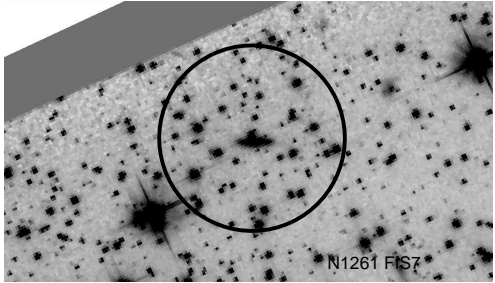}
  \end{center}
\caption{
HST/ACS preview image in F606W around N1261 FIS7, which is indicated
by a red circle. The north is up and the east is left,
and the field of view is $40\arcsec \times 20\arcsec$.
\label{fig:N1261FIS7}}
\end{minipage}
\end{figure}

Here, we further argue that our sources are contaminated
with background galaxies.
NGC~1261 was included in the FIRBACK survey at $170~\micron$,
performed with the ISO \citep{Puget-1999}.
Four sources we detected in the WIDE-S image were already
found in their survey, namely, FSM\verb|_|001 (FIS2), FSM\verb|_|002
(FIS3), FSM\verb|_|003 (FIS6), and FSM\verb|_|007 (FIS8).  
\citet{Patris-2003} made spectroscopic 
follow-up of the FIRBACK sources and found that the four sources
have redshifts between 0.03 and 0.13. We also found galaxies for
two other FIS sources with the aid of SIMBAD
database,\footnote{This research has made use of the SIMBAD database,
operated at CDS, Strasbourg,
France.\\http://simbad.u-strasbg.fr/simbad/}
MZZ~10046 (emission-line galaxy) for FIS1 and
ESO-LV~155-0100 for FIS4.
For FIS7, we found a counterpart of galaxy in the preview image
of HST/ACS\footnote{Based on observations made with the NASA/ESA
Hubble Space Telescope, obtained from the data archive at
the Space Telescope Science Institute. STScI is operated by
the Association of Universities for Research in Astronomy,
Inc. under NASA contract NAS 5-26555.} in both F606W and F814W
filters (figure \ref{fig:N1261FIS7}).
The object is elongated, which suggests that the object is a galaxy.
That leaves us only FIS5 that has no
hint of background galaxy identification.
It is located out of the HST/ACS image.
However, its SED looks just like that of the FIS7.
It is likely that FIS5 is also a background galaxy.
Although the field of NGC~1261 contains the largest number of
detected sources
among our survey, none of them seems to be 
associated with the intracluster dust.

For other clusters, there is not much information about the natures
of the point-like sources. They do not gather around the cluster
center.
If they are associated with
the clusters, the number detected in each cluster
is expected to depend on distance to the cluster
because of the limited sensitivity.
Such a trend, however, was not found.
On the other hand, the number of confusing galactic sources
should be dependent on galactic latitude, which is not the case again.
We suggest that most of the FIS sources
are extragalactic sources, probably background star-forming galaxies
based on their similar SEDs to the sources toward NGC1261.

\subsection{Remaining candidate of the intracluster dust}
\label{sec:candidate}

\begin{figure}
\begin{minipage}{80mm}
  \begin{center}
    \FigureFile(70mm,90mm){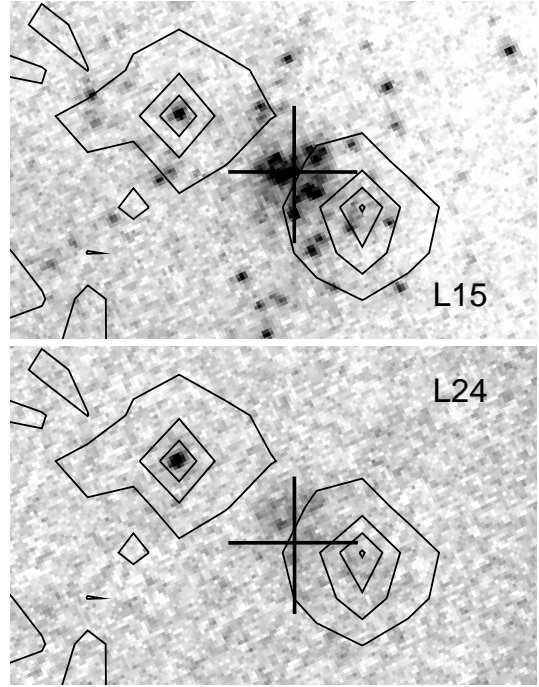}
  \end{center}
\caption{
The close-up figures of the IRC (L15, L24) images 
of NGC~5024. The central coordinate of the cluster
is indicated by the cross.
The north is up and the east is left, and the fields-of-view
about $6\arcmin \times 3.5\arcmin$.
The contours indicate the flux density
in the WIDE-S filter. Among two peaks,
the west one corresponds to N5024 FIS2, and the east one to FIS3.
\label{fig:NGC5024cen}}
\end{minipage}
\end{figure}

We argue that N5024 FIS2 is a candidate of the intracluster dust.
First, there is no mid-infrared counterpart found in the IRC images,
disclaiming that it is a galaxy containing star forming regions
within. In figure \ref{fig:NGC5024cen}, we present that 
there is no counterpart of
N5024 FIS2, which is in contrast to a neighboring source N5024 FIS3.
Although there is a few mid-infrared sources around the peak
of N5024 FIS2, none of them is prominent in the L24 image.
Compared with the SEDs in figure \ref{fig:FISIRCSEDs},
N5024 FIS2 has smaller mid-infrared emission.

\begin{figure}
\begin{minipage}{80mm}
  \begin{center}
    \FigureFile(70mm,60mm){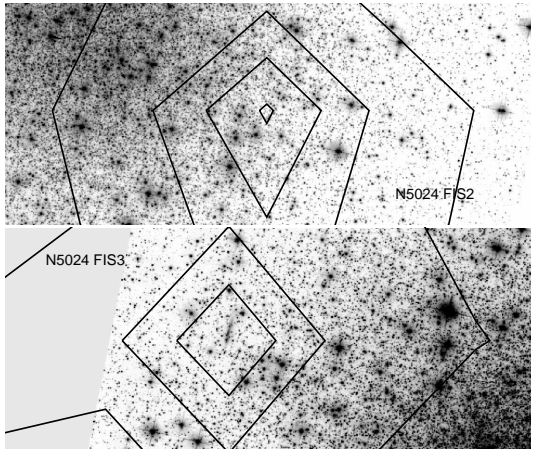}
  \end{center}
\caption{
The Close-up of an HST/ACS preview image 
in F606W
at around N5024 FIS2 and FIS3. The north is up and the east is left,
and the fields-of-view are $110\arcsec \times 75\arcsec$.
The contours indicate the flux density common 
with those in figure \ref{fig:NGC5024cen}.
\label{fig:NGC5024HST}}
\end{minipage}
\end{figure}

Figure \ref{fig:NGC5024HST} presents
the HST/ACS preview images in F606W around the coordinates of
the FIS sources, N5024 FIS2 and FIS3.
While there is a background galaxy found at
the peak of N5024 FIS3, we found no apparent galaxy for the FIS2.
We confirmed the same result in the F814W image.
This supports the intracluster origin
although it is possible that a fainter galaxy
or a quasar is associated with them.
It is necessary to carry out observations in other wavelength
to conclude its nature (see section \ref{sec:future}).

\section{Discussion}\label{sec:Discussion}

\subsection{Upper limit of the dust mass from
the background fluctuation}
\label{sec:MassLimit}

In this section, we assess the upper limit of undetected
intracluster dust within the half-mass radius $r_{\mathrm{h}}$
based on the background fluctuation.
We follow the equation of \citet{Lynch-1990},
\begin{equation}
f_\nu = 6 \sqrt{N} A_{\mathrm{p}} \sigma _\nu,
\end{equation}
where $N$ is the number of pixels corresponding to the area,
$A_{\mathrm{p}}$ is
the area of the pixel, and $\sigma _\nu$ is the standard deviation
of the background, to estimate the flux limits (eq. \ref{eq:Mdust}).
We list the limits of $90~\micron$ flux in table \ref{tab:limits}
with some parameters of the clusters.

According to \citet{Boyer-2006}, the flux $f_\nu$ [mJy]
can be converted into the dust mass as,
\begin{equation}
M_{\mathrm{d}} = 4.79 \times 10 ^{-17} ~f_\nu
\frac{D_{\mathrm{sun}} ^2}{\kappa _\nu B_\nu (T_{\mathrm{d}})}
~[{\rm M}_\odot],
\label{eq:Mdust}
\end{equation}
under the assumption that the dust cloud is optically thin.
$D_{\mathrm{sun}}$ is the distance in kilo-parsec,
$\kappa _\nu$ is the dust absorption coefficient in cm$^2$~g$^{-1}$, 
$B_\nu (T_{\mathrm{d}})$  is the Planck function in cgs units, and
$T_{\mathrm{d}}$ is the dust temperature.
We calculate $\kappa _\nu = 40$ cm$^2$~g$^{-1}$ 
from \citet{Ossenkopf-1994} assuming a standard
Mathis-Rumpl-Nordsieck dust distribution
as \citet{Boyer-2006} did.
The temperature has a significant effect on the estimate of the 
dust
mass through the Planck function $B _\nu (T _{\mathrm{d}})$.
At $90~\micron$, its dependency on the temperature is as
$B_{\mathrm{100K}}:B_{\mathrm{70K}}:B_{\mathrm{35K}} = 2.2:1:0.09$,
which directly changes the mass estimate according to
equation (\ref{eq:Mdust}).
\citet{Angeletti-1982} performed realistic estimates
of the temperature of intracluster dust, and obtained
temperatures between 35 and 90~K dependent on the assumed chemical
composition of dust grain.
Here we use two assumed dust temperatures
of 35~K and 70~K. The value of 70~K is adopted
from the temperature of M~15 IR1a\citep{Boyer-2006}.
The $M_{\mathrm{d}}$
gets smaller in case of $T_{\mathrm{d}}=90$~K,
which accentuates the lack of intracluster dust.
The results are listed in table \ref{tab:limits}.

\begin{table*}
  \begin{minipage}{0.99\hsize}
  \caption{
  Parameters of the observed globular clusters and obtained upper limits.
  The updated version of \citet{Harris-1996} is used to take
  metallicities [Fe/H], integrated magnitudes $\mathrm{M_V}$, distances
  $D_{\mathrm{sun}}$, and heights $Z$ above or below the galactic plane.
  Escape velocities at half-mass radii are taken from \citet{Gnedin-2002}.
  If available, the velocities $V_{\mathrm{Z}}$ perpendicular to the plane 
  are taken from \citet{Dinescu-1999} and \citet{Casetti-Dinescu-2007}.
  The $3 \sigma$ upper limits of
  the emission $f_{90\micron}$ and the mass $M_{\mathrm{d}}$
  of the intracluster dust are obtained within the half-mass
  radii $r_{\mathrm{h}}$, while
  $\tau _{\mathrm{d}}$ is the upper limit of the lifetime of
  the dust obtained by equation (\ref{eq:lifetime}).
  Two values are listed based on the two assumed dust temperatures,
  $T_{\mathrm{d}}=35 and 70$~K, for both $M_{\mathrm{d}}$ and
  $\tau _{\mathrm{d}}$.
  The $M_{\mathrm{d}}$ values for N5024 FIS2 (this work) and for M~15
  and NGC~6356 (from early papers) are also listed and
  their lifetimes are obtained in the same manner.
  \label{tab:limits}}
  \begin{center}
    \begin{tabular}{crrrrrrrrrrr}
      \hline
Object  & \multicolumn{1}{c}{[Fe/H]} &
   \multicolumn{1}{c}{M$_{\mathrm{V}}$} &
   \multicolumn{1}{c}{$V_{\mathrm{esc}}^{\mathrm{h}}$} &
   \multicolumn{1}{c}{$D_{\mathrm{sun}}$} &
   \multicolumn{1}{c}{$Z$} &
   \multicolumn{1}{c}{$V_{\mathrm{Z}}$} &
   \multicolumn{1}{c}{$f_{90\micron}$} &
   \multicolumn{2}{c}{$M_{\mathrm{d}}$} &
   \multicolumn{2}{c}{$\tau _{\mathrm{d}}$} \\
& \multicolumn{1}{c}{dex} & \multicolumn{1}{c}{mag} &
\multicolumn{1}{c}{km/s} & \multicolumn{1}{c}{kpc} &
\multicolumn{1}{c}{kpc} & \multicolumn{1}{c}{km/s} &
\multicolumn{1}{c}{mJy} & \multicolumn{2}{c}{$10^{-5}$~M$_\odot$} &
\multicolumn{2}{c}{$10^6$~yr} \\
& & & & & & & & 70~K & 35~K & 70~K & 35~K \\ 
\hline
 NGC1261 & $-1.35$ & $ -7.81$ & 20.2 & 16.4 & 12.9 &  ---   &  16.9 &  8.8 & 96 &  1.9 & 20 \\
 NGC1851 & $-1.22$ & $ -8.33$ & 24.3 & 12.1 &  6.9 & $-109$ &  10.6 &  3.0 & 32 &  0.3 &  3 \\
 NGC1904 & $-1.57$ & $ -7.86$ & 26.1 & 12.9 &  6.3 & $   6$ &  14.5 &  4.6 & 50 &  1.6 & 17 \\
 NGC2808 & $-1.15$ & $ -9.39$ & 45.8 &  9.6 &  1.9 & $  63$ &  25.8 &  4.6 & 50 &  0.1 &  2 \\
 NGC5024$^*$ & $-1.99$ & $ -8.70$ & 20.9 & 17.8 & 17.5 & $ -83$ &  20.1 & 12.3 &133 &  5.0 & 55 \\
 NGC5139 & $-1.62$ & $-10.29$ & 44.0 &  5.3 &  1.4 & $  -9$ & 122.8 &  6.7 & 72 &  0.3 &  3 \\
 NGC5272 & $-1.57$ & $ -8.93$ & 22.7 & 10.4 & 10.2 & $-154$ &  20.2 &  4.2 & 46 &  0.5 &  6 \\
 NGC5634 & $-1.88$ & $ -7.69$ & 17.0 & 25.2 & 19.1 &  ---   &  11.0 & 13.5 &146 & 10.9 &118 \\
 NGC5904 & $-1.27$ & $ -8.81$ & 29.2 &  7.5 &  5.4 & $-222$ &  33.4 &  3.6 & 39 &  0.3 &  3 \\
 NGC6205 & $-1.54$ & $ -8.70$ & 26.9 &  7.7 &  5.0 & $-119$ &  26.9 &  3.1 & 34 &  0.4 &  5 \\
 NGC6341 & $-2.28$ & $ -8.20$ & 29.2 &  8.2 &  4.7 & $  35$ &  19.7 &  2.6 & 28 &  3.2 & 35 \\
 NGC6402 & $-1.39$ & $ -9.12$ & 26.2 &  9.3 &  2.4 &  ---   &  72.9 & 12.2 &132 &  0.9 &  9 \\
\hline
N5024 FIS2 & $-1.99$ & $ -8.70$ & 20.9 & 17.8 & 17.5 & 1.11 &
   \multicolumn{1}{c}{114} & 50 & 500 &  10 & 100 \\
NGC~7078 & $-2.26$ & $-9.17$ & 27.4 & 10.3 & 4.7 & 1.06 &
   \multicolumn{1}{c}{---} &
   90$^\dagger$ & \multicolumn{1}{c}{---} &
   20 & \multicolumn{1}{c}{---}  \\
NGC~6356 & $-0.50$ & $-8.52$ & 31.7 & 15.2 & 2.7 & 0.74 &
   \multicolumn{1}{c}{---} & 860$^\ddagger$ & --- & 
    7 & --- \\
\hline
    \end{tabular}
  \end{center}
  $^*$In the event that N5024 FIS2 is not associated
     with the cluster.\\
  $^\dagger$\citet{Boyer-2006},~
  $^\ddagger$\citet{Hopwood-1998}.
  \end{minipage}
\end{table*}

\begin{figure}
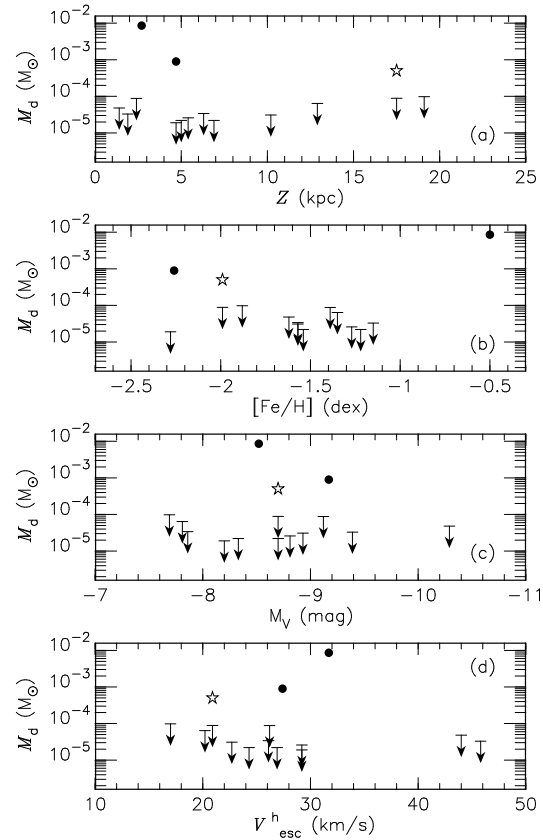

\begin{minipage}{80mm}
  \begin{center}
    \FigureFile(70mm,105mm){fig11.ps}
  \end{center}
\caption{
The mass and upper limits of the intracluster dust plotted against
(a) $Z$-height, (b) metallicity [Fe/H], (c) integrated magnitude
$M_{\mathrm{V}}$, and (d) escape velocity at the half-mass radius
$V_{\mathrm{esc}}^{\mathrm{h}}$.
The $y$-axis value is based
on the assumption of $T_{\mathrm{d}}=70$~K.
In case of $T_{\mathrm{d}}=35$~K, all the points shift upward 
by a factor of 11. The arrows indicate
the upper limits obtained in this work,
and star symbols indicate the estimated mass of the intracluster-dust
candidate N5024 FIS2.
Two filled circles indicate M~15 (NGC~7078)
and NGC~6356 based on the estimate by \citet{Boyer-2006} and
\citet{Hopwood-1998} respectively.
\label{fig:Mdust}}
\end{minipage}
\end{figure}

In figure \ref{fig:Mdust}, we plot the estimated upper limits
of the intracluster dust against some parameters
of the clusters which are taken from \citet{Harris-1996} and
\citet{Gnedin-2002}\footnote{We used their machine-readable
tables on Web:\\
http://www.physics.mcmaster.ca/Globular.html\\
http://www.astro.lsa.umich.edu/\%7Eognedin/gc/vesc.dat}.
We used the Harris catalog available online which was updated in 2003.
We also include the mass of the intracluster-dust candidate,
N5024 FIS2 (see section \ref{sec:N5024FIS2}), and those of the dust clouds
in M~15 and NGC~6356 reported in previous studies.
No parameter can be an exclusive factor to determine
the accumulated mass of the intracluster dust.
Although \citet{Hopwood-1999} suggested that
the metallicity may be an influencing factor based
on the fewer samples with shallower observations,
the detection of M~15 IR1a disclaims such possibility.
It is unclear what parameter(s) determine the amount of
the accumulated intracluster dust.

The estimated
dust
mass of M~15 IRa1 is $(9 \pm 2) \times 10^{-4}$~M$_\odot$,
much larger than the upper limits in table \ref{tab:limits}.
\citet{Boyer-2006} used a box aperture of
$130\arcsec.5 \times 130\arcsec.5$,
which is larger than a circle with $r_{\mathrm{h}}=1.06\arcmin$.
Nonetheless, the size of the far-infrared emission of the detected
intracluster dust, IR1a, is smaller than the circle so that we can
make a direct comparison of the values for IR1a with
the upper limits obtained in this work.
It is clear that the intracluster dust in M~15, which is still
smaller than the estimated mass released from red giants
\citep{Boyer-2006},
is exceptionally large compared with possible intracluster dust in
other globular clusters if any.
The parameters such as $V_{\mathrm{esc}}^{\mathrm{h}}$
and $Z$-height for at least some of our samples are comparable with
those of M~15.

\subsection{The lack of intracluster dust}\label{sec:LackDust}

Here we discuss the lack of intracluster dust quantitatively
by estimating the lifetime from the expected amount of gas
released from red giants.
We can calculate the gas amount corresponding to that of
the intracluster dust with the dust-to-gas mass ratio $\psi$.
The ratio is considered to be dependent on metallicity.
\citet{Marshall-2004} concluded that $\psi$ is proportional to
the metallicity by comparing OH/IR stars in the solar neighborhood
and the large Magellanic Cloud (also see \cite{vanLoon-2006}).
It is worthwhile noting uncertainties on $\psi$ especially
in the low-metallicity environment.
\citet{Marshall-2004} found
the proportionality between $\psi$ and metallicity
based on the objects between $\sim 0.1~Z_\odot$ and $\sim 1~Z_\odot$.
It is not clear if the extrapolation can be applied below
[Fe/H]~$= -1$~dex.
\citet{Boyer-2006} found more than 20 stars with dusty wind in M15.
Although they did not give any estimate on the dust mass-loss rate,
some of them have large mid-infrared excesses comparable with
the mass-losing stars in the metal-rich cluster 47~Tuc
\citep{Origlia-2007}.
Such dusty mass-loss was not expected in a metal-poor cluster
like M~15.
If $\psi$ is not so small as expected from the proportionality,
the amount of gas corresponding to the upper limits of dust
gets smaller.
The $\psi$ value can be changed by
non-canonical evolution of stars such as interacting binaries
and/or blue stragglers.
In the following estimations, we simply assume
$\psi = 0.01 \times 10^{\mathrm{[Fe/H]}}$.

It is possible to evaluate the number of mass-losing stars
in a cluster by using the specific evolutionary flux,
\begin{equation}
B \sim 2 \times 10^{-11} {\rm ~stars ~yr^{-1} ~L_\odot^{-1}},
\end{equation}
which is the number of stars entering or leaving
any post-main-sequence stage per year and per solar luminosity of
the stellar population \citep{Renzini-1986}. The average dust mass
released per year within a cluster can be obtained as,
\begin{equation}
F = \Delta M_{\mathrm{d}} B L_{\mathrm{cl}},
\label{eq:MassLoss}
\end{equation}
where $\Delta M_{\mathrm{d}}$ is the amount of dust to be lost
during the post-main-sequence stage and $L_{\mathrm{cl}}$ is
the total luminosity of the cluster. We assume that each red giant
in a cluster loses
the gas of 0.2~M$_\odot$, then $\Delta M_{\mathrm{d}}$ is
calculated as $0.2 \psi$~M$_\odot$
$= 2 \times 10^{-3+{\mathrm{[Fe/H]}}}$~M$_\odot$.
We obtain
the upper limits of the lifetime of the intracluster dust:
\begin{equation}
\tau _{\mathrm{d}} = \frac{M_{\mathrm{d}}}{F} =
   \frac{M_{\mathrm{d}}}{\Delta M_{\mathrm{d}} B L_{\mathrm{cl}}/2}.
\label{eq:lifetime}
\end{equation}
$L_{\mathrm{cl}}$ is divided by two because the upper limit
$M_{\mathrm{d}}$ has been obtained for the intracluster dust within
the half-mass radius. This factor may not be correct if the dust
released from stars outside
the half-mass radius falls into the central region of the cluster,
in which case the estimated lifetime $\tau_{\mathrm{d}}$ gets even
smaller.

We list the obtained lifetimes in table \ref{tab:limits}.
Two values are obtained based on the assumed dust temperatures
of 35~K and 70~K. The value with $T_{\mathrm{d}}=35$~K
is 11 times larger than the one with $T_{\mathrm{d}}=70$~K
from the ratio of Planck function. In both cases,
they are significantly shorter than the typical time interval,
$\sim 10^8$~yr, between passages through the galactic plane
for most of the clusters.
Please note that the $\tau _{\mathrm{d}}$ gets even smaller
if $\psi$ is larger than the value expected from 
the proportionality between $\psi$ and [Fe/H].

The kinematics information of the clusters suggests that
most of the clusters have actually spent
a significant time as large as $10^8$~yr since the last passage.
Here we consider the orbits obtained by \citet{Dinescu-1999}
and \citet{Casetti-Dinescu-2007}.
The velocity components perpendicular to the galactic plane,
i.e. $V_{\mathrm{Z}}$ listed in table \ref{tab:limits},
indicate that five clusters are now approaching to
the plane. For NGC~1904 and 5139, $V_{\mathrm{Z}}$ almost 
equals to zero, so that they are in the middle of the passage
through the galactic plane.
Time since the most recent passages is about
the half of the orbital periods or longer
for these seven clusters.
NGC~1851 and 6341 are still moving away from the plane.
NGC~1851 is well separated from the plane by 6.9~kpc.
The obtained lifetime, $\tau_{\mathrm{d}} < 3$~Myr,
is certainly shorter than the time since its last passage.
For NGC~6341, on the other hand, the discrepancy
between the obtained lifetime and the time since
the last passage is not large to conclude the lack 
of the intracluster dust in case of $T_{\mathrm{d}}=35$~K.
There are no proper motions available for NGC~1261, 5634,
and 6402.

We confirmed that the estimated lifetimes are shorter
than the duration in which the intracluster dust can be
accumulated since the cluster crossed the galactic plane
for at least seven clusters in our sample.
The conclusion for NGC~5024 depends on whether
N5024 FIS2 belongs to the cluster or not.
The short lifetimes indicate that
the intracluster dust is not accumulated as was expected
and disappears in a short time scale. 
Thanks to the high sensitivity of the AKARI/FIS,
we could conclude the lack of the intracluster dust
even with the lower dust temperature of 35~K.

\subsection{Estimated mass of the intracluster-dust candidate}
\label{sec:N5024FIS2}

We have found a candidate of the intracluster dust,
N5024 FIS2. Here, we estimate its mass assuming that it is
associated with the cluster.
The diffuse emission around 
NGC~6402 can be associated with the cluster as discussed
in section \ref{sec:NGC6402}. However, the clumpy background
prevents us from a robust mass estimation, so that we do not
include it in quantitative discussions below.

The temperature for N5024 FIS2 cannot
be determined well based on our data because of the small coverage
of wavelength (only in N60 and WIDE-S) and
the limited photometric accuracy ($\pm~20$\%).
The flux ratio of $F_{90}/F_{65} = 1.33\pm 0.4$ corresponds to
$T=55\pm 15$~[K]. We perform two estimates
assuming 70~K and 35~K again. Then, the dust
mass of N5024 FIS2 is estimated at $5 \times 10^{-4}$~M$_\odot$
or $5 \times 10^{-3}$~M$_\odot$ for 70~K and 35~K respectively
from equation (\ref{eq:Mdust}). The former
value agrees within a factor of 2 with that of M~15 IR1a
\citep{Boyer-2006}.
Under the assumption of the proportionality between $\psi$ and [Fe/H],
the corresponding gas mass
is 5~M$_\odot$ (or 50~M$_\odot$ if $T_{\mathrm{d}}=35$~K)
so that it 
is made up of the dust released from a number of stars.
The obtained lifetime $10^7$~yr (or $10^8$~yr)
is longer than the upper limits obtained above. 
If we assume the same dust temperature as the cases of
M~15 and NGC~6356,
$\tau_{\mathrm{d}}$ for N5024 FIS2 is comparable
with the lifetimes
for the intracluster dust in these clusters.

There remain a few features to be explained if N5024 FIS2 is actually
a dust cloud  within the cluster. It is offset from
the cluster center, i.e. the bottom of the gravitational potential,
by $52\arcsec$ $(\sim 0.8 r_{\mathrm{h}})$.
M~15 IR1a is located about $17\arcsec$ $(\sim 0.3 r_{\mathrm{h}})$
to the west of
the cluster core. \citet{Boyer-2006} argued that the intracluster dust
is short-lived and has not been dynamically relaxed.
This can be true for N5024 FIS2 too.
Another intriguing feature for our case is that N5024 FIS2 looks like
a point source.
The estimated mass suggests that a couple of
dozen mass-losing stars contributed to N5024 FIS2.
The dust clouds released from mass-losing stars
aggregated into one blob and it is constrained in a rather small
region providing N5024 FIS2 is associated with the cluster.
On the other hand, if the dust-to-gas ratio $\psi$ does not
follow the proportionality (see section \ref{sec:LackDust}),
the estimated dust mass of N5024 FIS2 does not necessarily exceed
the amount released from one object. With $\psi = 0.01$
and $T_{\mathrm{d}}=70$~K assumed, the estimated amount
of the corresponding gas reduces to 0.05~M$_\odot$.
The size evolution of ejecta from a single AGB star is discussed
by \citet{Wareing-2007}. In the interaction with
the interstellar medium, the size of circumstellar shell
can be restricted to within a few pc, which is as small as
the FWHM ($\sim 40\arcsec$) of the FIS at the distance
of NGC~5024 (17.8~kpc).
It is essential to explain these features in order to conclude
the intracluster-dust origin of N5024 FIS2.
Theoretical work may be useful to study
how the materials released from red giants evolve in
the intracluster space and interact with each other.

\subsection{
Impacts on evolution of globular cluster
}

As mentioned in the Introduction, intracluster matter
may have played an important role in
the chemical evolution of globular clusters.
\citet{Tsujimoto-2007} calculated that
a significant amount of the mass released
from intermediate-mass AGB stars can be accreted to surrounding stars
in NGC~5139 during 2~Gyr. However, it is possible that
no significant mass remains to be accreted
if the intracluster matter disappears
with the short lifetime mentioned above.
Of course, it is not dust but gas that should be considered as
the source of the mass to be accreted because
\citet{Tsujimoto-2007} are interested in the enrichment of helium.
The fate of the intracluster gas can be rather different from
that of the intracluster dust. Previous observations,
however, suggest that the intracluster gas is also
deficient in globular clusters as mentioned in the introduction.
The deficiency was also reported for NGC~5139 \citep{Smith-1990}.
The conclusion of \citet{Tsujimoto-2007} can be
altered if the intracluster gas is removed
before it is accreted to the stars within the cluster.
Likewise, the fate of the intracluster dust can have
an effect on the possible chemical enrichment via the ejecta
of evolved stars especially for heavy elements depleted in dust.
For example, aluminum abundance pattern has been also explained
by the ejecta of AGB stars (\cite{Denissenkov-1998},
\cite{Ventura-2008}).

In any case,
it is of interest to investigate the evolution,
destruction, and/or removal
of the intracluster gas and dust in the context of their effects
on the evolution of globular clusters.
It is possible that globular clusters were embedded
in dark matter halos once in the early universe
and the intracluster-matter evolution was different
from that in current clusters
\citep{Bekki-2006}.
Nonetheless, observation of current ones
can provide important constraints 
to be compared with detailed modeling.

\subsection{Future prospects}\label{sec:future}

Besides pointed observations as those presented here,
AKARI accomplished
the All-Sky Survey. The expected sensitively is lower than
those of the pointed observations. The 5~$\sigma$ detection limit 
in the WIDE-S filter is 0.55 Jy for a point source
\citep{Kawada-2007}.
Nonetheless, it is similar to that in 
previous observations made with the ISO which resulted in
a couple of detections (\cite{Hopwood-1998}; \cite{Evans-2003}).
The All-Sky Survey data will enable us to search for
the intracluster dust
in all the globular clusters with a reasonable detection limit.

On the other hand, with the high sensitivity of the FIS pointed
observations,
a significant number of background galaxies appear
above the detection limit. In the future surveys,
it can become a problem to discriminate the intracluster dust
from the background galaxies unless they are more extended than
expected for galaxies. A potentially useful method is
to combine the radio observation for the sources.
It is well known that the FIR-radio correlation holds
for various kinds of galaxies \citep{Dickey-1984}.
\citet{Oyabu-2005} explored the correlation for the far-infrared
sources in the Lockman Hole region based on the ISO's survey
(90 and $170~\micron$)
and 1.4~GHz observations with Very large Array (VLA). We can obtain
$\log (F_{90\micron} / F_{\mathrm{1.4GHz}}) \sim 2.8$ based on
their data. This value increases
to about 6.2 if we consider the blackbody radiation of 70~K. 
Therefore, the radio flux clearly discriminates between
the intracluster dust and background galaxies assuming
that intracluster dust has a temperature of 70~K as M~15 IR1a.
As for the source we detected, $F_{90\micron}$ of about 100 mJy
predicts $F_{\mathrm{1.4GHz}} \sim 700$ $\mu$Jy for the galaxies and
$\sim 60$~nano-Jy for the blackbody radiation of 70~K.
Though the expected $F_{\mathrm{1.4GHz}}$ value is lower than
the detection limits of the wide-field survey catalog
currently available even for the case of galaxies,
deep follow-up observations may tell which is the case.

\section{Summary}\label{sec:Summary}

We obtained far-infrared images for 12 globular clusters
with the AKARI/FIS, which allowed us to do deep surveys
of the cold intracluster dust. While we found one candidate
of the intracluster dust,
we confirmed the paucity of such dust in most clusters.

In the FIS images, we detected extended emissions
toward two clusters, namely NGC~2808 and NGC~6402.
They are separated from the galactic plane by less than
15 degrees, and large-scale emission of the galactic cirrus
is dominant in their fields. We suggest that
the detected bumps of emission are merely
fluctuations of the background emission.
In case of NGC~2808, the emission is largely away
from the cluster center, while the bump
coincides with the center of NGC~6402. We cannot
rule out the possible dust component associated with
NGC~6402.

On the other hand,
we detected 28 point-like sources, but
concluded that most of them are background galaxies.
For those with the near-to-mid infrared data by the AKARI/IRC,
the obtained SEDs resemble those of galaxies.
We found one source, N5024 FIS2, without any clear counterpart
in the mid infrared. There is no galaxy found in the HST/ACS image
at around the coordinate. 
It is a possible candidate of the intracluster dust.
If so and the temperature of 70~K is assumed, the
dust
mass is estimated at $5 \times 10^{-4}$~M$_\odot$. However,
some intriguing features such as the point-like appearance
remain to be explained if N5024 FIS2
actually belongs to the cluster.
Besides the candidate, our results give upper limits
of 2--9 $\times 10^{-5}$~M$_\odot$ within the half-mass radii of
the 12 clusters from the background fluctuation.

Some mechanism(s) are clearly efficient to remove
the intracluster dust in most clusters, if not all.
The lifetime of the intracluster dust
is as short as $10^{5-6}$ yr.
On the other hand, we found that the intracluster dust
detected in M~15 and NGC~6356 is exceptionally large compared with
other clusters. Our result strengthens the mystery of
the lack of the intracluster dust.
It urges us to study physical processes which determine the fate
of the intracluster dust. Together with that of the intracluster gas,
it has a potential to affect the evolution of
globular clusters.

\bigskip

We acknowledges the anonymous referee whose helpful comments
gave us a chance to improve the paper.
We thank Shinki Oyabu for his suggestions about the usefulness of
radio follow-up observation.
We are grateful to Iain McDonald for 
helpful comments based on
his analysis on the Spitzer Space Telescope data.

\end{document}